\documentclass[]{spie}  %>>> use for US letter paper
\usepackage[]{graphicx}
\def\etal{{\em et al.}}

\newcommand{\be}{\begin{equation}}
\newcommand{\ee}{\end{equation}}
\newcommand{\ba}{\begin{eqnarray}}
\newcommand{\ea}{\end{eqnarray}}
\newcommand{\ban}{\begin{eqnarray*}}
\newcommand{\ean}{\end{eqnarray*}}
\newcommand{\baa}{\begin{array}}
\newcommand{\eaa}{\end{array}}

\def\micron{$\mu$m}

\def\gev{GeV~c$^{-2}$}

\def\cm2{cm$^2$}

\def\perm2{m$^{-2}$}

\def\percm3{cm$^{-3}$}

\def\gevpercm3{\gev~\percm3}
\def\gpercm3{g~\percm3}

\def\calif252{$^{252}$Cf}
\def\pb210{$^{210}$Pb}
\def\bi210{$^{210}$Bi}
\def\po210{$^{210}$Po}
\def\am241{$^{241}$Am}
\def\k40{$^{40}$K}
\def\c14{$^{14}$C}
\def\co60{$^{60}$Co}

\def\h0{H_{0}}

\def\g300{G_{300}}

\def\r350{R_{350}}

\def\q300{Q_{300}}

\title{ARCONS: A highly multiplexed superconducting optical to near-IR camera } 

\author{Benjamin A. Mazin\supit{a}, Kieran O'Brien\supit{a}, Sean McHugh\supit{a}, Bruce Bumble\supit{b}, David Moore\supit{c}, Sunil Golwala\supit{c}, Jonas Zmuidzinas\supit{c}
\skiplinehalf
\supit{a}Department of Physics, University of California, Santa Barbara, CA, USA; \\
\supit{b}NASA Jet Propulsion Laboratory, 4800 Oak Grove Drive, Pasadena, CA, USA; \\
\supit{c}Department of Physics, California Institute of Technology, 1200 E. California Blvd., Pasadena, CA, USA \\
}

\authorinfo{Further author information: (Send correspondence to B.A.M.)\\B.A.M.: E-mail: bmazin@physics.ucsb.edu, Telephone: (805)893-3344}
\begin{document} 
  \maketitle 

%%%%%%%%%%%%%%%%%%%%%%%%%%%%%%%%%%%%%%%%%%%%%%%%%%%%%%%%%%%%% 
\begin{abstract}
We report on the development of ARCONS, the ARray Camera for Optical to Near-IR Spectrophotometry.  This photon counting integral field unit (IFU), being built at UCSB and Caltech with detectors fabricated at JPL, will use a unique, highly multiplexed low temperature detector technology known as Microwave Kinetic Inductance Detectors (MKIDs).  These detectors, which operate at 100 mK, should provide photon counting with energy resolution of $R=E/\delta E>20$ and time resolution of a microsecond, with a quantum efficiency of around 50\%. We expect to field the instrument at the Palomar 200" telescope in the first quarter of 2011 with an array containing 1024 pixels in a 32$\times$32 pixel form factor to yield a field of view of approximately 10$\times$10 arcseconds.  The bandwidth of the camera is limited by the rising sky count rate at longer wavelengths, but we anticipate a bandwidth of 0.35 to 1.35~\micron~will be achievable.  A simple optical path and compact dewar utilizing a cryogen-free adiabatic demagnetization refridgerator (ADR) allows the camera to be deployed quickly at Naysmith or Coud\'{e} foci at a variety of telescopes.  A highly expandable software defined radio (SDR) readout that can scale up to much larger arrays has been developed.
\end{abstract}

%>>>> Include a list of keywords after the abstract 

\keywords{MKIDs, ARCONS, cryogenic detectors, IFU, LTD, kinetic inductance detector}

%%%%%%%%%%%%%%%%%%%%%%%%%%%%%%%%%%%%%%%%%%%%%%%%%%%%%%%%%%%%%
\section{INTRODUCTION}
\label{sec:intro}  % \label{} allows reference to this section

Over the last three decades improvements in our detectors, and the instruments and telescopes that feed them, has led to a golden age in astronomy and ushered in the era of precision cosmology.  While the fundamental detector technologies of the past 30 years have been based on semiconductors, new detectors that use superconductors to significantly increase performance are currently being developed for wavelengths from the radio to the X-ray~\cite{Wilson:2000p48,Romani:2001p1716,Martin:2004p1697}.  These detectors operate below 1 Kelvin to reduce the contributions of thermal noise, allowing them to measure the energy and arrival time of a single photon.  A collaboration led by UCSB is building a camera based on a highly multiplexible type of low temperature detectors (LTDs) known as Microwave Kinetic Inductance Detector (MKID)~\cite{Day:2003p180,Mazin:2006p104,Mazin:2008p59} optimized optical and near-infrared astronomy. The goal of ARCONS, the ARray Camera for Optical to Near-IR Spectrophotometry, is to both demonstrate the viability of the MKID technology as well as do science that would be difficult or impossible with conventional instruments.

The camera will use Optical Lumped Element (OLE) MKIDs~\cite{Doyle:2008p278} designed at UCSB and fabricated at JPL.  These detectors are based on a TiN film with a controllable superconducting transition temperature, allowing tuning of the device during fabrication for the maximum energy resolution for a given operating temperature.  The first generation MKIDs, based on tantalum absorbers, have a maximum theoretical energy resolution, $R=E/\delta E=50$ at 400 nm~\cite{Mazin:2004p2}.  The second generation MKIDs used in this investigation have a theoretical energy resolution at 400 nm of $R=200$ for an operating temperature of 50 mK, and $R=400$ for an operating temperature of 15 mK, eight times higher than the first generation MKIDs.  These devices are also not spatially multiplexed into strip detectors~\cite{Mazin:2006p104}, which allows a maximum count rate 20 times higher than the first generation MKIDs.  This higher count rate will allow us to extend the simultaneous spectral coverage, limited by the rising sky background count rate in the near infrared, from the 350--750 nm of the first generation detectors to 350--1350 nm.  This detector technology allow ARCONS to have significant advantages over a conventional lenslet, fiber fed, or image slicer integral field unit (IFU):

\begin{itemize}
\item Time resolution up to six orders of magnitude better than a CCD
\item Extremely broad intrinsic bandwidth, with good quantum efficiency from 0.1--6 $\mu$m; nearly 10 times the bandwidth of a CCD
\item No read noise or dark current, and nearly perfect cosmic ray rejection
\item Scalable to much larger arrays than conventional IFUs 
\item Time domain information allows the use of a calibration star for dynamic apertures and software tip/tilt corrections, greatly improving signal to noise and allowing the instrument to take immediate advantage of improved seeing conditions
\item Photon arrival time, energy resolution, and the large number of pixels will allow us to monitor and remove sky emission, allowing Poisson-limited operation even in spectral regions dominated by OH emission 
\end{itemize}

Two limitations are that the energy resolution, while ideal for faint objects, is not high enough for some applications, and the array size is small compared to an imaging CCD (although not small compared to IFUs).  The first problem can be solved by using the detectors as the order sorter for an echelle spectrograph, although this feature will not be included in ARCONS.  The size of the arrays are currently only limited by what we can afford to readout with our room temperature readouts (Section~\ref{sec:readout}), which are improving with Moore's Law.  Megapixel arrays are possible within a decade.

\section{MICROWAVE KINETIC INDUCTANCE DETECTORS}
\label{sec:mkid}

\begin{figure}
   \begin{center}
   \begin{tabular}{c}
   \includegraphics[width=1.0\columnwidth]{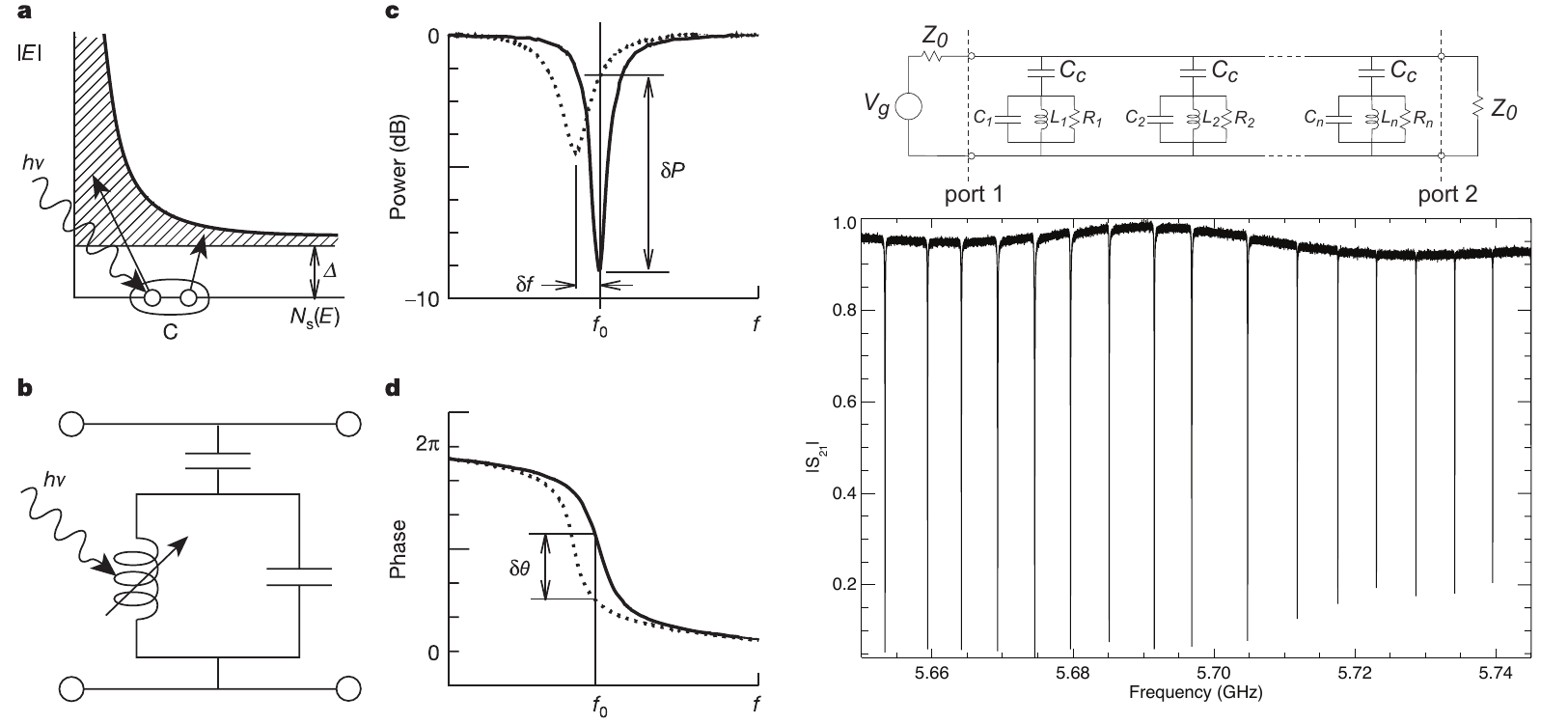}
   \end{tabular}
   \end{center}
   \caption 
   { \label{fig:detcartoon} 
Left: An illustration of the operational principle behind a MKID.  Right: An array of CPW MKIDs fabricated with a 40 nm aluminum film on sapphire show a mean frequency jitter of 0.8 MHz.  This demonstrates that FDM multiplexing with a frequency spacing less than 2 MHz is feasible.}
   \end{figure}

Microwave Kinetic Inductance Detectors (MKIDs) are a relatively new and promising type of superconducting
photon detector~\cite{Day:2003p180,Mazin:2006p104,Mazin:2008p59}. The primary advantage of MKIDs compared to other low temperature
detectors is that by using resonant circuits
with high quality factors, passive frequency domain multiplexing
allows up to thousands of resonators to be read out through a single
coaxial cable and a single high electron mobility transistor (HEMT)
amplifier (Figure~\ref{fig:detcartoon}, right). Large arrays of MKIDs are significantly easier to
fabricate and read out than any competing technology. They do not
require any superconducting electronics, and their readouts can
leverage the tremendous advances in room temperature microwave
integrated circuits developed for the wireless communications
industry~\cite{chervenak99,jonas04,mazin06}, as shown in Section~\ref{sec:readout}.

MKIDs work on the principle that incident photons change the surface
impedance of a superconductor through the kinetic inductance effect.  The kinetic inductance effect occurs because energy can be stored in the supercurrent of a superconductor.  Reversing the direction of the supercurrent requires extracting the kinetic energy stored in the supercurrent, which yields an extra inductance.
This change can be accurately measured using a thin film
superconducting resonant circuit, resulting in a measurement of the
energy and arrival time of the incident photon for the case of
optical/UV/X-ray photons or the total absorbed power for lower energy
photons.  The only real difference between arrays designed for
different wavelengths is the method used to couple the photon energy
into the MKID --- the detectors themselves and the readout are
nearly identical. Figure~\ref{fig:detcartoon} gives an overview of
this process. In Figure~\ref{fig:detcartoon} panel (a), a photon
with energy $h\nu > 2\Delta$ ($\Delta$ is the superconducting gap
energy) is absorbed in a superconducting film cooled to $T \ll T_c$,
breaking Cooper pairs and creating a number of quasiparticle
excitations $N_{qp} = \eta h\nu / \Delta$.  The efficiency of
creating quasiparticles $\eta$ will be less than one since some of
the energy of the photon will end up as vibrations in the lattice
called phonons. In this diagram, Cooper pairs (C) are shown at the
Fermi level, and the density of states for quasiparticles, $N_s(E)$,
is plotted as the shaded area as a function of quasiparticle energy
$E$.

Panel (b) shows that the increase in quasiparticle density changes
the surface impedance $Z_s = R_s + i \omega L_s$ of the film
(represented as the variable inductor), which is used as part of a
microwave resonant circuit. The resonant circuit is depicted
schematically here as a parallel $LC$ circuit which is capacitively
coupled to a through line. The effect of the surface inductance
$L_s$ is to increase the total inductance $L$, while the effect of
the surface resistance $R_s$ is to make the inductor slightly lossy
(adding a series resistance).

Panel (c) shows that on resonance, the $LC$ circuit loads the
through line, producing a dip in its transmission. The
quasiparticles produced by the photon increase both $L_s$ and $R_s$,
which moves the resonance to lower frequency (due to $L_s$) and
makes the dip broader and shallower (due to $R_s$). Both of these
effects contribute to changing the amplitude (c) and phase (d) of a
microwave probe signal transmitted past the circuit.  The amplitude and phase curves shown in this illustration are actually the data measured for an aluminum test device at 120~mK (solid lines) and
260~mK (dotted lines), which is of a magnitude similar to what would
be expected from a photon event. 

If the resonator is excited with a constant on-resonance microwave signal, the energy of the absorbed photon can be determined by measuring the amount of phase and amplitude shift. The phase and amplitude change quickly (several $\mu$s) during the initial absorption event, followed by a slow ($\sim$100 $\mu$s) return to the unexcited state as the quasiparticles recombine into Cooper Pairs. This produces a characteristic phase and amplitude pulse for each arriving photon.  A Wiener optimal filter can be used to extract the initial photon energy from this pulse data.

As can be seen from Figure~\ref{fig:detcartoon}c, the transmission through the resonator is very high for signals detuned from the resonance. Since the quality factor (the resonant frequency divided by the line width) of these resonators is very high, in the range of 10,000--3,000,000, the degree of detuning is very small before the transmission is perfect. This leads to a simple readout multiplexing scheme. Many resonators with slightly different resonance frequencies may be coupled to the same feedline, as shown on the right top panel of Figure~\ref{fig:detcartoon}. The array is excited with a microwave comb function containing a sine wave at the resonant frequency of each of the resonators in the array. The resonators pick out one particular excitation signal while passing all of the rest unaltered. The signals for the entire array can be amplified with a single cryogenic amplifier without saturation. We have fabricated arrays with 640 resonators, and have demonstrated multiplexed readout of a subset of the array with a mean resonator to resonator frequency jitter of 0.8 MHz using CPW MKIDs, as shown in the bottom right panel of Figure~\ref{fig:detcartoon}. It is possible to read out an array of thousands of detectors with this technique using only a single transmission line~\cite{mazin06}.

   \begin{figure}
   \begin{center}
   \begin{tabular}{c}
   \includegraphics[width=1.0\columnwidth]{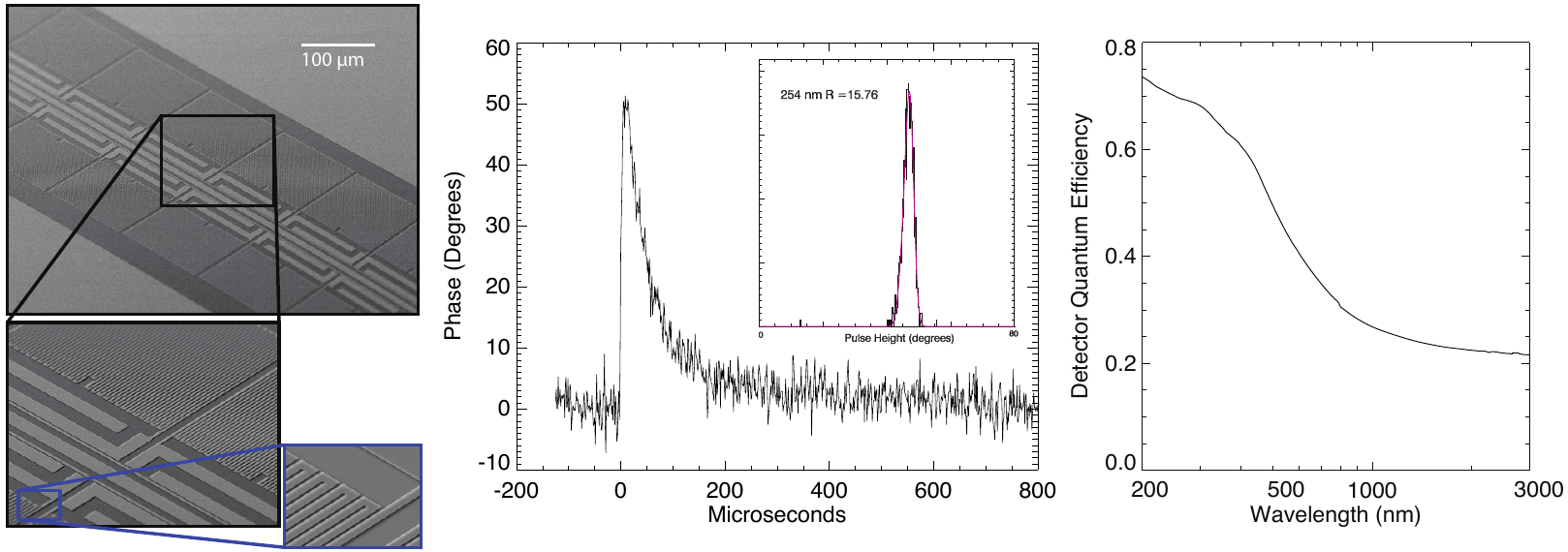}
   \end{tabular}
   \end{center}
   \caption
   { \label{fig:ole} 
Left: A scanning electron micrograph (SEM) of a prototype OLE MKID. Center: The typical phase response of a OLE MKID when illuminated with 254 nm photons.  The inset shows actual data from an OLE MKID, consisting of a histogram of 5000 photons with an energy resolution $R=E/\delta E=16$. Right: The measured absorption of a 40 nm thick TiN film on sapphire. This is roughly the detector quantum efficiency we can expect from the OLE MKID.}
   \end{figure}

Any resonator that uses a superconductor for its inductor can function as a MKID.  ARCONS will be using titanium nitride (TiN) optical lumped element (OLE) MKIDs~\cite{Doyle:2008p278}, formed by using an interdigitated capacitor attached to a inductive meander (Figure~\ref{fig:ole}, left) that functions as both the inductor and the photon absorber.  The advantage of this approach over strip detectors that use a separate absorber is that it eliminates quasiparticle trapping since the inductor also functions as the photon absorber, allowing the detectors to achieve the best theoretical energy resolution for a given operating temperature.  TiN was chosen as the superconductor because its superconducting transition temperature can be adjusted by varying the nitrogen content of the film, it has a very long penetration depth leading to resonators dominated by kinetic inductance, and it has been shown to form extremely high Q resonators~\cite{Leduc:2010p3489}.

The OLE design, shown in Figure~\ref{fig:ole}, is a simple one layer structure fabricated by optical lithography out of a sputtered TiN film on a sapphire or silicon wafer. The inductive meander of an OLE MKID is patterned with nearly the smallest feature size practical in order to increase the active area of the sensor. Our prototype OLE MKID uses 0.5~\micron~wide slots and 3~\micron~wires. This arrangement reduces detector quantum efficiency by 17\% at wavelengths shorter than 0.5~\micron, but provides the full quantum efficiency of the TiN film for wavelengths longer than 0.5~\micron~where the inductor looks like a uniform superconducting sheet to the incoming photons. Slot widths of 0.25~\micron~should be straightforward with modern lithographic techniques, removing this 17\% loss.  In addition, a microlens array will be employed to focus the incoming light only on the sensitive area of the resonator, allowing nearly 100\% fill factor if a square microlens array is used.  This allows a larger capacitor, which lowers the detector TLS noise.  A finite ground plane CPW feedline will pass near the capacitor, allowing us to tune the coupling quality factor $Q_c$ by changing the distance between the capacitor and the feedline.  The resonant frequency of each pixel will be set by slightly changing the length (and hence the inductance) of the inductor.

TiN has a gold color, and can be quite reflective in the infrared, as measured in the right panel of Figure~\ref{fig:ole}.  Since the quantum efficiency of the device is determined by the amount of light absorbed by the metal, we plan to use a surface treatment to improve the absorption.  We expect to achieve a quantum efficiency of 70\% over the entire 0.35--1.35 \micron~band.

\section{ARCONS}
\label{sec:ARCONS}

ARCONS is based around an adiabatic demagnetization refrigerator (ADR) with a base temperature of 70 mK.  It will contain 1024 pixels, making it the largest optical/UV LTD camera by an order of magnitude.  ARCONS will have a bandwidth of 350--1350 nm, limited by the sky count rate, and a field of view of approximately $10\times10$ arcseconds.  We expect an instrument throughput approaching $40\%$.  We plan on fielding ARCONS at the Palomar observatory in the first quarter of 2011.  The energy and time resolution combined with good system efficiency make this a powerful instrument.  

\subsection{Detector and Cooling}
\label{sec:detcool}

The heart of the instrument is a cryogen-free adiabatic demagnetization refrigerator (ADR) with optical access, as shown in the left panel of Figure~\ref{fig:ARCONS}.  This cryostat uses a two-stage pulse tube to reach 3 Kelvin, and a 4 Telsa superconducting magnet and two paramagnetic salt pills to reach a base temperature of 70 mK.  The hold time below 100 mK is 2 days with no excess heat load.

Two stainless steel coaxial cables bring microwave signals to and from the 3 K stage.  From the 3 K stage, a superconducting NbTi coaxial cable brings the signals to the array, and then another NbTi coax bring the signals back into a low noise cryogenic HEMT amplifier mounted on the 3 K stage.  The array is mounted in a microwave packaging designed to suppress box resonances.  The device assembly is surrounded by a magnetic shield to reduce any potential noise from varying magnetic fields.  

The difficulty in opening an optical path from the dewar to the outside world is that room temperature blackbody radiation could heat the detector and cold stage, increasing detector noise and reducing hold time.  In order to block thermal infrared most of the glass in the optical system is at a temperature of 3 K or lower.  In addition, a Lyot stop tied to the 3 K stage can be fitted with various glasses up to 1 cm thick and even a metallic thick grill filter to block thermal infrared if required.  The exact design of the filtering system will be iterated since infrared transmission data on the glasses in the optical system is difficult to find.  

\subsection{Optical Design}
\label{sec:optical}

\begin{figure}
\begin{center}
\includegraphics[width=1.0\columnwidth]{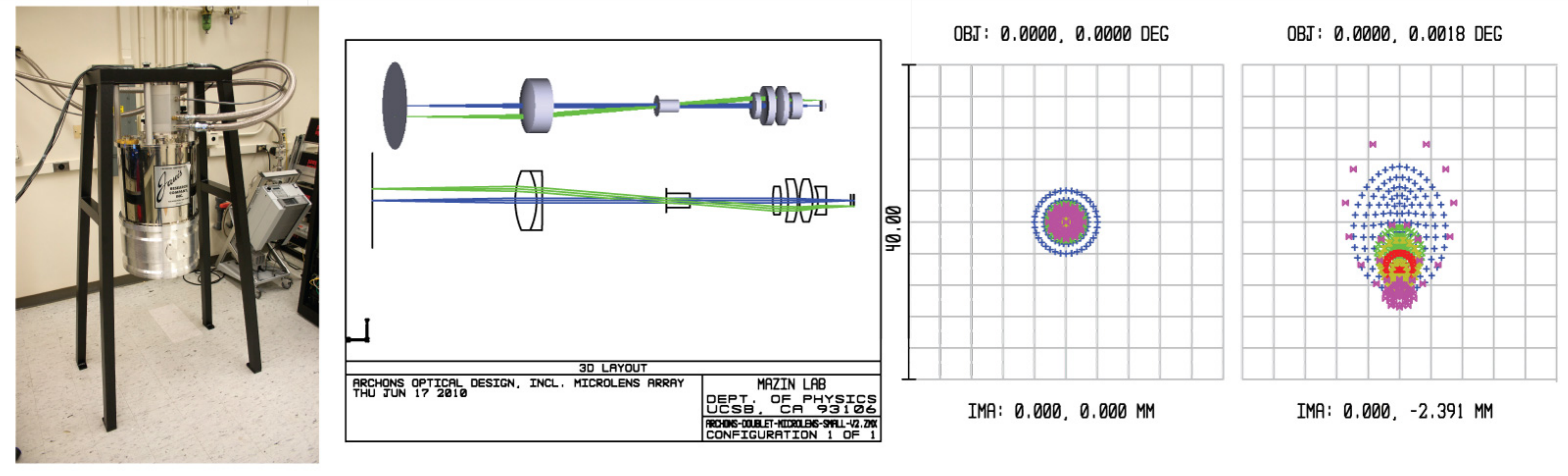}
\end{center}
\caption{Left: The cryostat that will house ARCONS. Center:  The optical train that will be used to couple ARCONS to the Palomar 200" Coud\'{e} focus. Right: A spot diagram showing that the performance of the optical system will be adequate across the entire field of view.} 
\label{fig:ARCONS}
\end{figure}

An optical system, shown in Figure~\ref{fig:ARCONS}, has been designed with Zemax to couple ARCONS with the Palomar Coud\'{e} focus.  This focus is stationary, which makes interfacing with the ADR (and its compressor) significantly easier, and the time resolution of the MKIDs means that rotation of the image plane over time can be easily removed.  

The dewar will be placed in the Coud\'{e} antechamber, and a flat mirror will direct the light into the cryostat.  The observatory guide camera will also be pointed at this oversized flat mirror.  The optical system consists of an external filter wheel containing several different shortpass filters to cut off the infrared, with edges at 750, 1000, and 1350 nm.  These filters will allow us to change the full instrument bandwidth depending on the atmospheric conditions and lunar phase.  After the filter wheel is a shutter to allow dark measurement of the array, followed by an achromatic doublet to collimate the beam, which also serves as the vacuum window of the dewar.  The next optical element is the 1 cm thick piece of glass at 3 K that forms the Lyot stop.  This glass will be selected to block any infrared transmission windows in the BK7 and SF6 glass that makes the cold optics.  If required, a thick grill filter will be made with lithography on this piece of glass to block out unwanted infrared.

After the Lyot stop, a physically separated achromatic doublet (to avoid problems with differential expansion) sets the plate scale, and two further optical elements reimage the focal plane.  The resulting spot diagram can be seen in the right panel of Figure~\ref{fig:ARCONS}.  The simulation includes the microlens array in front of the detector.  The MKIDs pixel pitch of 100~\micron~means that the optical system does not need high magnification, simplifying the system.  The system is designed to have 0.33" pixels on the sky.  The clear aperture of the optical system was designed to allow an unvignetted field of view of $20\times20$ arcseconds.  

\subsection{Digital Room Temperature Readout}
\label{sec:readout}

In order to read out an MKID array, one must generate a comb of frequencies with a sine wave at the resonant frequency of each individual resonator.  The frequency comb is then sent through the device, where each detector imprints a record of its illumination on its corresponding sine wave.  The comb is then amplified with a cryogenic HEMT and brought outside the cryostat, where it is then digitized and the phase and amplitude modulation of each individual sine wave is recovered in room temperature electronics.  Aside from a simple HEMT amplifier, there are no cryoelectronics.  Compared to existing TES SQUID multiplexers, much of the complexity is moved from the base temperature to room temperature, where the full power of modern microwave electronics is available.  This readout technique was first demonstrated for MKIDs in 2006~\cite{mazin06}.  

\begin{figure}
\begin{center}
\includegraphics[width=1.0\columnwidth]{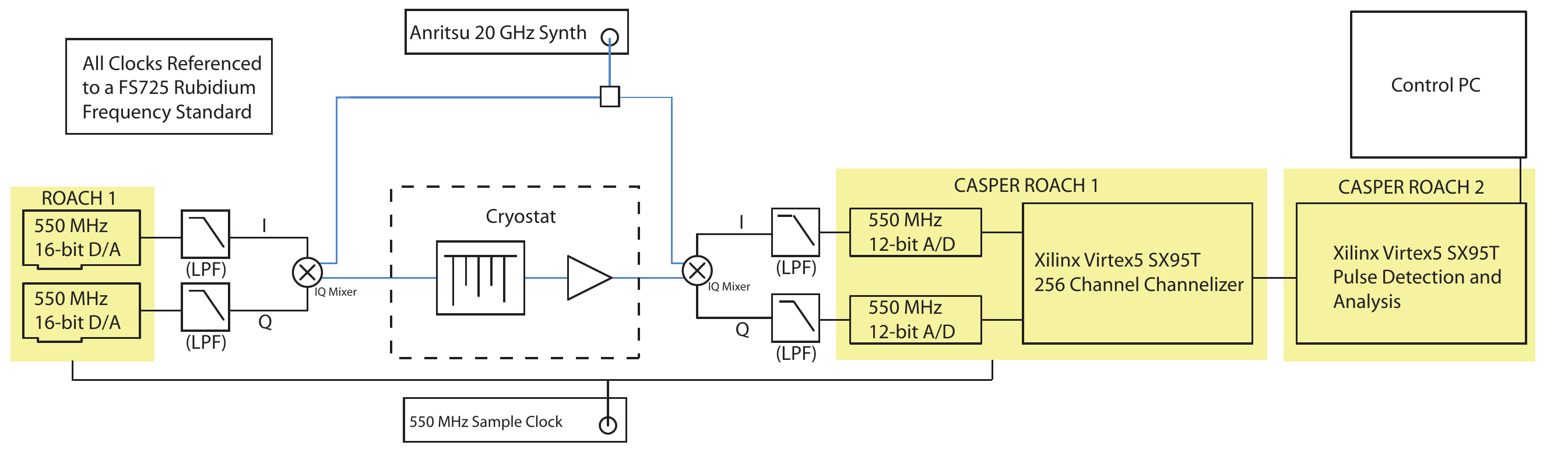}
\end{center}
\caption{A block diagram of the Open Source Resonator Readout currently under construction.} \label{fig:sdr}
\end{figure}

The technique described above, where a comb of frequencies is created, modified, then digitized and analyzed, is very common in modern wireless communications, where it is usually referred to as software-defined radio (SDR).  We have formed a collaboration of seven universities and national labs to build an open source resonator readout around the Berkeley CASPER group's hardware for ARCONS and other upcoming projects.  Our implementation is shown in Figure~\ref{fig:sdr}. In this implementation, dual 550 MHz, 16-bit digital to analog converters play back precomputed sine waves to generate the comb.  Since two D/As are used, we can use an IQ modulator, which allows us to produce signals within a 550 MHz wide band centered on our local oscillator (LO) frequency (usually 2--12 GHz).  After the comb passes through the detector, it is mixed back down to baseband with another IQ modulator, low-pass filtered, then digitized with dual 550 MHz, 12-bit analog to digital converters.  After digitization, the signals are passed to a fast field programmable gate array (FPGA).  There are many algorithms that can be run in the FPGA to demodulate, or channelize, the signals.  The simplest is a direct digital downconverter (DDC) that simply digitally multiplies the complex input signal by sine wave at the desired frequency.  This shifts the frequency of interest to 0 Hz.  A digital filter followed by decimation gives the desired output data stream.  We are using a more complex two stage channelization core that uses a polyphase filter bank (PFB) followed by a time-multiplexed direct digital downconverter to allow us to readout 256 resonators in 550 MHz of bandwidth.  The readout hardware is complete and operating in the lab, and the software and firmware is currently under development.  This readout will have final cost of about \$35 per resonator.  We expect this price to drop with Moore's Law, so in 10 years the cost should be around \$0.35 per resonator, easily allowing megapixel readouts.  

\section{SCIENCE}
\label{sec:science}

In the last several years groups have begun to build instruments
based around integral field units (IFUs), using either optical
fibers, lenslets, or image slicers to rearrange the light hitting the focal
plane. Conventional integral field spectrographs can have higher
energy resolution than MKID focal planes, but they also have
significantly lower end-to-end efficiency, limited bandwidth, and can be limited by read
noise since the spectra are dispersed over many pixels. Our low
resolution, high efficiency approach will enable science on objects
too faint for conventional high resolution IFUs, as well as adding a unique high time resolution capability.
There are a large class of general astrophysical problems that can
be addressed with these arrays.  If the targets are rare enough
that there is only one in the telescope field of view, having a
small focal plane array imposes no penalty.  The 10"$\times$10"
field of view of ARCONS will allow us to sometimes get bright (20$^{th}$ magnitude) calibrators on the array,
which will allow software tip/tilt corrections and the use of
dynamic apertures that have been shown to increase the signal to
noise ratio of observations~\cite{shearer97}.  While the complexity of the instrument and its data products requires it to remain a private instrument, the investigators will actively seek science collaborators to help maximize the scientific output of this unique instrument.

Listed below are three projects we plan to pursue with ARCONS.  This is just a small sampling of the science ARCONS will enable, since many observation of rare, very faint objects could benefit from this technology.

\subsection{Optical Pulsars and Magnetars}

Pulsars remain an enigma despite over 40 years of intense study.  In principle, they are a laboratory where scientists can study electromagnetism, gravity, and nuclear physics at fields strengths and densities that will likely forever remain outside the realm of laboratory experiment.  Despite intense study in the radio and gamma-ray, the location and mechanism of the optical through gamma-ray emission in pulsars remains a mystery.  For example, polar cap, outer gap, two-pole caustic, and slot-gap models~\cite{Cheng:2000p2100,kern03,dyks04,Harding:2008p1282} do not readily explain the current data on optical emission.  Untangling the emission mechanism in the optical should lead to a better overall understanding of pulsar emission at all
wavelengths.

ARCONS will be a powerful tool to help unravel this mystery.  Simultaneous optical, radio, and gamma ray observations provide a means of testing models of high-energy emission.  The investigators will produce spectrally-resolved light curves of pulsars, to obtain greater sensitivity to features that change with pulse phase, and to track their variation. Polarization of optical emission is expected and is understandable as a consequence of the pulsar’s magnetic field; simple polarization optics will allow measurement of polarization in the later years of the investigation. We plan to compare optical data with simultaneous gamma-ray and radio data.  Gamma-ray data are publicly available from Fermi, and simultaneous radio observations are easily arranged, particularly for stronger sources.  Searches for optical variability, mirroring efforts in the radio and in the gamma-ray with Fermi, may also provide insight into the emission mechanism.  

\begin{figure}
\begin{center}
\includegraphics[width=1.00\columnwidth]{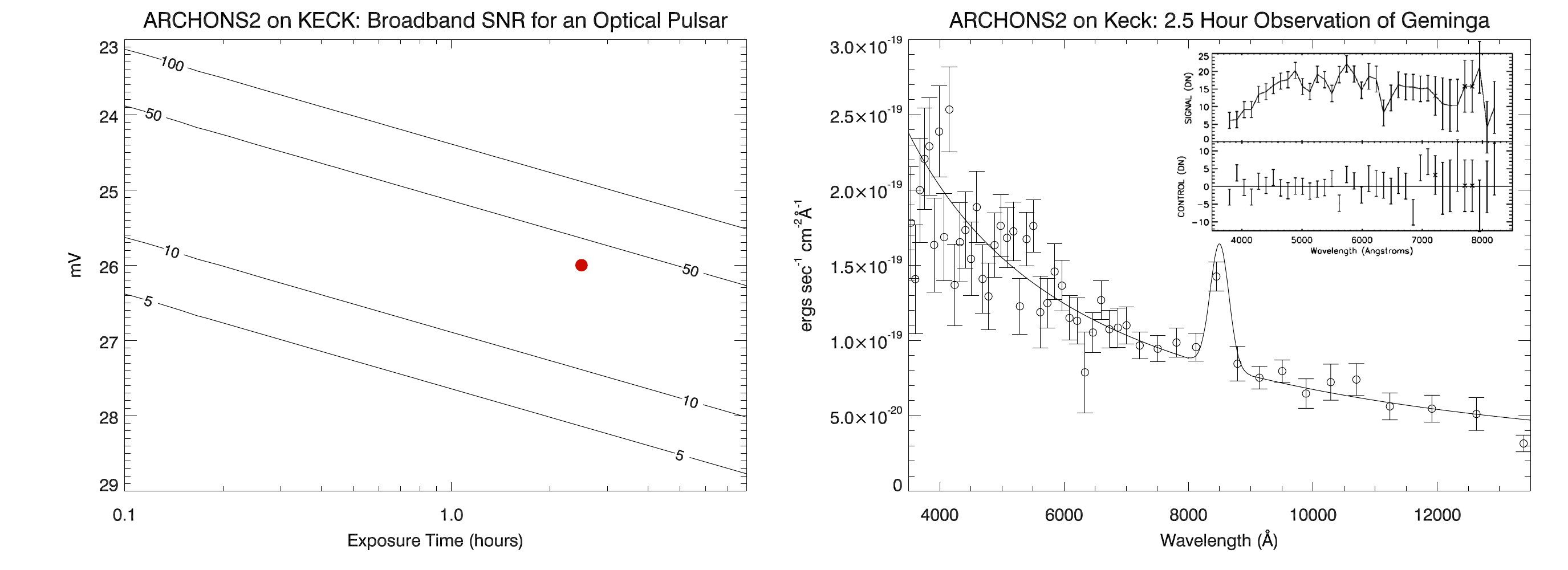}
\end{center}
\caption{Left: Simulation of the broadband signal to noise (SNR) ratio of an optical pulsar with a Geminga-like power law spectrum $(f_\nu=\nu^{-0.8})$~\cite{Martin:1998p2146} observed with ARCONS on Keck in 0.8" seeing with an energy resolution R=50.  The red dot represents the 2.5 hour simulated observation of Geminga shown in the right panel.  Right: A 2.5 hour simulated observation of Geminga showing the clear detection of a hypothetical cyclotron emission line at 850 nm with a strength of twice the continuum and a FWHM of 35 nm.  The black line shows the model spectra used to simulate the observation.  The inset shows an the actual spectra of Geminga taken in a 2.5 hour observation with LRIS on Keck by Martin \etal.~\cite{Martin:1998p2146}.  The ARCONS spectra is superior mainly because of the ability to take advantage of the 0.8" second seeing to get a lower sky count rate and better control of systematics, such as variations in the intensity and spectrum of the night sky over the 30 minute LRIS exposures.} 
\label{fig:geminga}
\end{figure}

The first pulsar science targets with ARCONS will be known optical pulsars and magnetars like PSR B0656+14~\cite{Kern:2003p56}, Geminga, and 4U 0142. These objects are very faint, with V magnitudes around 25--27.  Observations will also focus on detecting optical pulsations in systems like PSR B1929+10 and PSR B0950+08, which have likely optical counterparts that have not yet been seen to pulse.  Another fruitful path will be attempt to detect optical counterparts to pulsars found in the Fermi catalog~\cite{abdo09}.  A simulated observation showing the power of ARCONS on Keck for low resolution spectroscopy is shown in Figure~\ref{fig:geminga}.  In an actual observation of a pulsar the spectra could be displayed as a image cube, allowing us to study the time variation of the spectra with changing pulse phase~\cite{Romani:2001p1716}. 

\subsection{X-ray Binaries}
With the advent of large telescopes, novel fast instrumentation and powerful X-ray satellites, we are beginning to unravel the correlated variability in Low Mass X-Ray Binaries (LMXBs). This is revealing a richer, more complex picture than first thought. While much of the optical/IR emission arises from the reprocessing of X-rays, a number of recent observations~\cite{Gandhi:2008p2158,Durant:2008p2211} have shown that there is also a component from the very rapid X-ray flickering on timescales of 10--150 ms, which is well correlated with optical/IR variability.

The reprocessed optical emission seen by a distant observer is delayed in time-of-arrival relative to the X-rays due to light travel times of order several seconds between the X-ray source and the reprocessing sites within the binary system. The time-delayed and distorted optical echoes directly measure the locations and sizes of reprocessing sites. By careful analysis of simultaneous optical and X-ray lightcurves, we recover not just the mean time delay, but also the range of time delays present between the two. The result of this ``deconvolution'' is a time-delay transfer function. By measuring the time-delay transfer function between the X-ray and the optical emission from the secondary (donor) star at two or more orbital phases, we can determine the inclination of the system. The absolute value of the delay, combined with the orbital phase of the observation, allows us to strongly constrain the combined masses of the two components in the binary from the orbital separation. We expect echo mapping to resolve LMXBs with a resolution of order 0.1-1 light seconds or better.

In LMXBs the size of the system is light seconds. In order to determine the time-delay transfer function, we need to study the correlated variability in X-ray and optical lightcurves. Several simultaneous X-ray/optical bursts have been observed to originate from 4U 1636-536 and EXO 0748-676. However, these observations used broadband optical photometry and were therefore relatively insensitive to the location of the reprocessed emission in the binary.  Time resolved spectrophotometry allows us to determine the spectrum of the time delays and search for correlations with spectral lines.

\subsection{Galaxies Formation and Evolution}

For extended objects, ARCONS on Keck will essentially act as a $32\times32$ pixel IFU with the sensitivity of LRIS and broader wavelength coverage. Since every pixel in ARCONS not pointed at an object is essentially a sky line monitor with microsecond time resolution and spectral resolution, we expect to be able to compensate for changes in both the intensity and spectrum of the sky lines extremely well, allowing us to be limited by Poisson noise even in spectral regions dominated by sky lines.  This will help in detecting extended emission and low surface brightness objects that would not be practical with a long slit spectrograph or conventional IFU.

ARCONS will be able to find redshifts of galaxies and quasars using spectral shape and Lyman-$\alpha$~emission~\cite{Mazin:2000p338,deBruijne02}. Figure~\ref{fig:geminga} shows the simulated spectra of a mV=26 object with a 2.5 hour integration time.  The wide wavelength range will allow us to see Lyman breaks at redshifts of $>$10. Imaging spectrophotometric data will determine not just the redshift~\cite{mazin00} of each object, but allow studies of the variations in stellar age, metallicity, and extinction across these faint sources.  

We suggest one initial extragalactic observational projects to demonstrate the power of ARCONS for extragalactic science, but there are many, many more interesting projects we will pursue with collaborators.  The unprecedented wavelength coverage of ARCONS will enable us to detect Lyman-$\alpha$ to a redshift of 10, up from the $\sim$6 with most current surveys.  Redshift 6--10 is an incredibly interesting time, as it is likely the era of reionization.  We will use ARCONS to follow up y and z band dropouts, such as those found by Bouwens \etal~\cite{Bouwens:2008p2309,Bouwens:2009p2329}. Looking for low surface brightness knots, emissions lines, and extended Lyman-$\alpha$ emission in these dropouts will confirm their redshift and allow us to study in detail some of the first galaxies.  

\section{CONCLUSIONS}
\label{sec:end}

When it is deployed at the Palomar 200" telescope in the first quarter of 2011, ARCONS will be the first camera using the MKID technology in the optical to near-IR.  It is the only single photon counting, energy resolving low temperature technology with a demonstrated multiplexing scheme that can be expanded to very large array sizes.  MKIDs are a transformational technology, and ARCONS will be the forerunner of a new generation of scientific instruments where the maximum amount of information can be extracted from every incoming photon.   

\acknowledgments     %>>>> equivalent to \section*{ACKNOWLEDGMENTS}       
 
This material is based upon work supported by the National Aeronautics and Space Administration under Grant NNH06ZDA001N-APRA2 issued through the Science Mission Directorate.  The authors would like to thank Rick LeDuc, Jiansong Gao, and David Pappas for useful insights.

%%%%%%%%%%%%%%%%%%%%%%%%%%%%%%%%%%%%%%%%%%%%%%%%%%%%%%%%%%%%%
%%%%% References %%%%%

\bibliography{/Users/bmazin/Data/Projects/bib/mazin.bib,/Users/bmazin/Data/Projects/bib/merged.bib}
\bibliographystyle{spiebib}   %>>>> makes bibtex use spiebib.bst

\end{document}